# Effect of electric field on the photoluminescence of polymer-inorganic nanoparticles composites

A. N. Aleshin*, I. P. Shcherbakov, E. L. Alexandrova, E. A. Lebedev

*Ioffe Physical-Technical Institute, Russian Academy of Sciences, St.Petersburg, 194021, Russia*

**Abstract**

We report on the effect of electric field on the photoluminescence (PL) from a composite consisting of a BEHP-*co*-MEH-PPV conjugated polymer mixed with ZnO nanoparticles. We have found that in the absence of electric field PL emission from the BEHP-*co*-MEH-PPV-ZnO composite has two maxima in the blue and green-yellow regions. Application of a voltage bias to planar gold electrodes suppresses the green-yellow emission and shifts the only PL emission maximum towards the blue region. Current-voltage characteristics of the BEHP-*co*-MEH-PPV-ZnO composite exhibit the non-linear behavior typical of non-homogeneous polymer-inorganic structures. Generation of excited states in the BEHP-*co*-MEH-PPV-ZnO structure implies the presence of several radiative recombination mechanisms including formation of polymer-nanoparticle complexes including "exciplex" states and charge transfer between the polymer and nanoparticle that can be controlled by an electric field.

---

## 1. Introduction

Embedding semiconducting nanoparticles into polymer matrices is an area of current interest in organic nanoelectronics. Such an integration of organic and inorganic materials of the nanometer scale into hybrid optoelectronic structures allows designing devices that combine the diversity and processibility of organic materials with high electronic and optical performance of inorganic nanocrystals. As a result of such integration one can obtain an increase in the efficiency of conventional organic light emitting diodes (OLEDs) as well as the possibility to tune the emission color by combining an efficient emission from both materials involved [1,2]. Intensive studies of luminescence properties of semiconducting polymers and some their composites with inorganic nanoparticles were carried out recently [3,4]. Because of the quantum size effect the color of emission can be varied by changing the nanocrystal size [5]. The most recent findings have revealed that it is possible to get the multicolor electroluminescent (EL) and photoluminescent (PL) emission from the same composite layer by switching the emission color with an applied electric field [2,6]. The electric field controlled switching of EL/PL emission between green and red color has been observed in composites based on poly(4,4'-diphenylene diphenylvinylene) (PDPV) and inorganic nanoparticles (complexes), such as ZnO:Mg, Ru. [2,6].

* Correspondant author, e-mail: aleshin@transport.ioffe.ru

However, such electric field controlled tuning of emission color in the blue-green region has not been demonstrated so far. Understanding the dominant electron-hole capture mechanism at the polymer-nanoparticle heterostructures is crucial for device optimization and has hence been discussed extensively in the literature [1-6]. At the same time, it is not clear till now whether capture takes place via tunneling and thermal injection or via barrier-free capture into the "exciplex" state [7,8]. To clarify all these questions, we have studied co-polymer - poly{[2-[2′,5′-bis(2″-ethylhexyloxy)phenyl]-1,4-phenylenevinylene]-co-[2-methoxy-5-(2′-ethylhexyloxy)-1,4-phenylenevinylene]}, (BEHP-co-MEH-PPV, m:n = 0.6:0.4) - an attractive conjugated polymer for EL/PL emission in the green region because of its high PL efficiency [3,9]. The chemical structure of BEHP-co-MEH-PPV ($E_g$ = 2.4 eV) is shown in Fig. 1a. The oxide semiconductor ZnO ($E_g$ = 3.35 eV, electron affinity = 2.7 eV) is a promising material for EL/PL emission in the blue-UV region [10]. The luminescence efficiency of ZnO nanoparticles can be even higher than that of bulk ZnO and the particle-size-dependent increase in the electronic band gap allows control of the excitation spectrum. The composite made of these two materials has not yet been investigated thoroughly that makes an analysis of the dominant mechanism of charge capture and radiative recombination in the BEHP-co-MEH-PPV-ZnO structure important.

In this paper we report on the effect of an electric field on transport and PL properties of the BEHP-co-MEH-PPV-ZnO composite that exhibits non-linear current-voltage characteristics (I-Vs) behavior typical of non-homogeneous polymer-inorganic structures and two PL maxima in the blue and green-yellow regions. Application of a voltage bias to planar gold electrodes suppresses the green-yellow emission and shifts the only PL emission maximum towards the blue region. Generation of excited states in the BEHP-co-MEH-PPV-ZnO structure implies the presence of several recombination mechanisms including formation of polymer-nanoparticle complexes including "exciplex" states and charge transfer between the polymer and nanoparticles that can be controlled by an electric field.

## 2. Experimental

The conjugated co-polymer - BEHP-co-MEH-PPV (m:n = 0.6:0.4) and ZnO nanoparticles (50-70 nm in diameter) used in our study were purchased from Sigma-Aldrich and used as received. BEHP-co-MEH-PPV was dissolved in chloroform (purity - 99.4%) with a concentration of 2-5 mg/ml. ZnO powder was dispersed in chloroform separately and subjected to ultrasonic treatment for 10 min. Then the BEHP-co-MEH-PPV solution and ZnO suspension were mixed together and subjected to ultrasonic treatment for 5-10 min. The mixture of BEHP-co-MEH-PPV and ZnO particles with different concentration ratios of components was drop-cast onto a Si substrate with a 200 nm thick $SiO_2$ layer on top and thermally evaporated gold electrodes (with thin Mo under layers to increase an adhesion). The distance between the electrodes was 15-30 μm, and the electrode width was 1 mm. The film thickness was ~ 0.4-0.6 μm. The samples were dried and heated at 80 C in $N_2$ atmosphere for 15 min and then kept in vacuum. The same procedure was used to prepare BEHP-co-MEH-PPV and ZnO films separately. The morphology of composite films was studied by an optical microscope with a CCD camera – Leica DM 2500M. A LGI-21 pulse laser ($\lambda$ = 337.1 nm, $E_i$ > $10^{-4}$ $J/cm^2$, $\tau$ ~ $10^{-8}$ s) was used to excite PL. The PL excitation beam was focused onto the area between the electrodes (see inset to Fig. 2a). The PL spectra were analyzed by using a SPM-2 spectrometer (spectral resolution ~ 2 nm) and a photo detector with a spectral sensitivity range 300-850 nm. I-Vs were studied in vacuum at room temperature by using a sample holder of a liquid $N_2$ cryostat and a dc electronic measuring system with a Keithley 6487 picoammeter/voltage source.

## 3. Results and Discussion

In order to understand the dominant mechanisms of charge capture and radiative recombination, we have studied the morphology, I-Vs, and PL spectra of the pristine BEHP-co-MEH-PPV film and ZnO particles as well as BEHP-co-MEH-PPV-ZnO composite films with different polymer/nanoparticles concentration ratios (1:1, 1:2, 1:4 etc.). Fig. 1a shows the chemical structure of the BEHP-co-MEH-PPV co-polymer, and Fig. 1b demonstrates the morphology of the BEHP-co-MEH-PPV-ZnO composite film (concentration ratio 1:4). As can be seen from Fig. 1b, the composite film is rather rugged, but ZnO particles are well dispersed in the BEHP-co-MEH-PPV matrix. Fig. 2a shows the PL spectra of the pristine ZnO particles (a) and the BEHP-co-MEH-PPV co-polymer film (b) used in our study. It is evident from Fig.2a that the BEHP-co-MEH-PPV film demonstrates well pronounced PL with the main emission maximum in the green-yellow region at $\lambda$ ~ 570 nm and weakly marked at $\lambda$ ~ 650 nm, which is consistent with the PL results obtained for this polymer earlier [9].

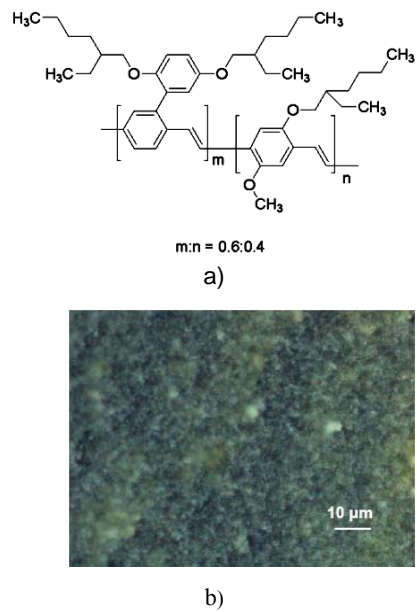

Fig. 1. Chemical structure of the BEHP-co-MEH-PPV co-polymer (a); morphology of the BEHP-co-MEH-PPV-ZnO composite film (b).

The PL spectrum of ZnO nanoparticles exhibits a strong PL maximum in the blue region ($\lambda$ ~ 390 nm), which corresponds well to the band gap energy of ZnO. The PL spectra of the BEHP-co-MEH-PPV-ZnO composite films with different polymer/nanoparticle concentration ratios (c - 1:1, d - 1:2, e - 1:4) are shown in Fig. 2b. As can be seen from Fig. 2b, the resulting spectra include the PL from both materials involved, which gives rise to two maxima in the blue (ZnO nanoparticles) and green-yellow (BEHP-co-MEH-PPV) regions. The comparison of the PL spectra of composites with various polymer/ZnO nanoparticles concentration ratios shows that in addition to the blue emission at 390 nm related to the ZnO particles, there are several new PL lines at 420, 460, 490 and 550 nm. The latter line is rather weak in the sample with a high polymer concentration (c) that has two additional lines at 570 and 650 nm which are likely to be related to the emission from co-polymer components. The increase in the ZnO nanoparticles concentration in the composite film leads to an efficient quenching of the green-yellow PL maxima at 570 and 650 nm attributed to the BEHP-co-MEH-PPV co-polymer. At the same time the intensity of the PL line at 550 nm increases. This leads to dramatic changes in the PL color from white (c) to light blue (e). As will be shown below, the PL line at 550 nm can be efficiently suppressed by applying an electric field due to separation of the charge carrier pairs related to the polymer-nanoparticle complexes.

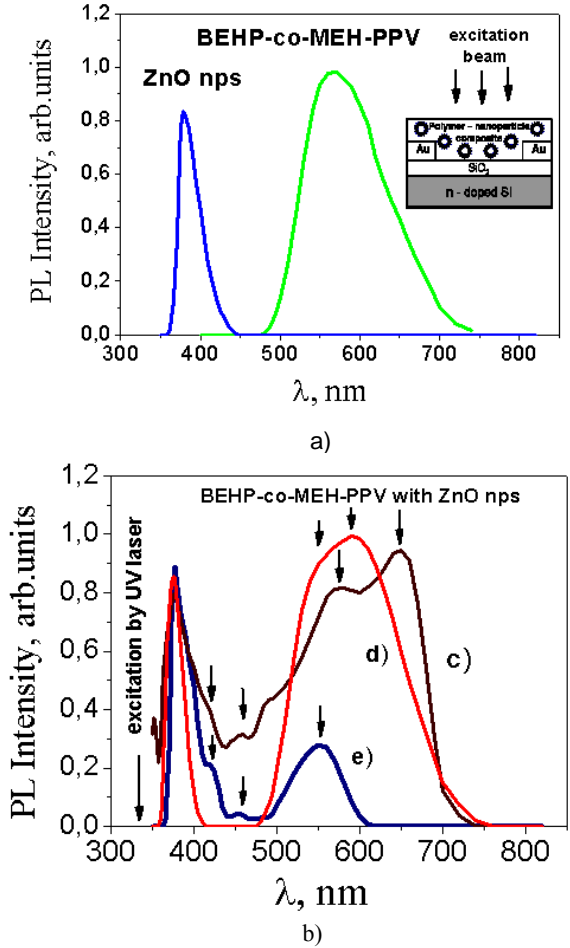

Fig. 2. a) PL spectra of the pristine BEHP-*co*-MEH-PPV co-polymer film (a) and ZnO particles (b) used in our study. Inset shows the composite film sample structure; b) PL spectra of the BEHP-*co*-MEH-PPV-ZnO composite films with different polymer/ZnO concentration ratios: c) 1:1, d) 1:2, e) 1:4.

Inset a) to Fig. 3 shows typical I-Vs of the BEHP-*co*-MEH-PPV-ZnO composite film (polymer/ZnO nanoparticles concentration ratio ~ 1:4) measured in a planar geometry (see inset to Fig. 2a). It is evident from this figure; I-Vs at 300 K are asymmetric and non-linear, which is characteristic of non-homogeneous polymer-inorganic heterostructures. Fig. 3 shows that lg-lg plot of these I-Vs follows the power law dependence:

$$I(V) \sim V^p \quad (1)$$

with three different power exponents (p) in different voltage ranges: 1) p ~ 0.9 - 1.4 (V < 1 V), 2) p ~ 2.5 - 3.0 (V~ 1 - 3 V), 3) 1.5 - 1.7 (V > 3 - 4 V). The shape of transient current for these composite films depends on the voltage bias (see Inset b) to Fig. 3). Namely, at V > 4 V there is a significant decrease in the current with time. The observed I-Vs behavior resembling that of thin PPV films caused by the space charge limited current mechanism (SCLC) [10, 11]. In the case of SCLC model, such a behavior of transient current can be associated with charge trapping on localized states. Therefore the observed I-Vs and transient currents are characteristic of charge injection into the composite film. Note that estimations of the current density in such composite structures results in values up to 1 A/cm$^2$ (at V ~ 10 V) which is much higher than that in conventional PPV films [12]. We suggest that such unusually high current densities in the Au-BEHP-co-MEH-PPV-ZnO-Au structure originate from the superposition of several transport mechanisms in composite layers, such as the SCLC with a possible double injection from Au contacts as well as percolation through polymer-ZnO nanoparticle interface network.

The influence of electric field on the PL emission from BEHP-*co*-MEH-PPV-ZnO composite films was studied by applying a voltage bias to planar electrodes when the PL excitation beam was focused onto the area between the electrodes as shown in the inset to Fig. 2a. As can be seen from Fig. 4, an application of a positive bias (electric field up to ~ 3 10$^4$ V/cm) to such an Au-BEHP-*co*-MEH-PPV-ZnO-Au structure suppressed significantly the green-yellow emission and shifted the emission color to the dark blue region related to ZnO nanoparticles (sample "e" in Fig.2b, polymer/ZnO nanoparticle concentration ratio ~ 1:4). The electric field controlled the PL emission shift from a bright to a dark blue color was found to be reversible. It was found that this effect is more pronounced for the composite samples with high concentration of ZnO nanoparticles.

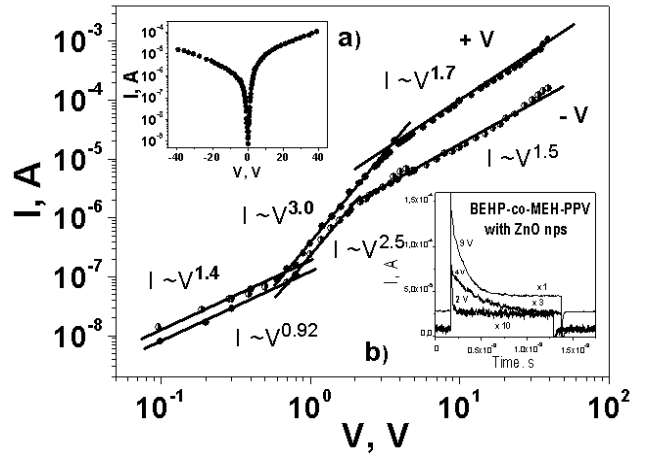

Fig. 3 Lg-lg plots of I-Vs of the BEHP-*co*-MEH-PPV-ZnO composite film at 300 K (concentration ratio ~ 1:4). Inset a) shows I-Vs of the same BEHP-*co*-MEH-PPV-ZnO composite film; inset b) demonstrates the transient current vs time dependencies for the BEHP-*co*-MEH-PPV-ZnO composite film at different voltage biases, V: 2 (x10); 4 (x3); 9 (x1).

Note that the effect of electric field on the emission color has also been observed in PDPV films doped with ruthenium dinuclear complex [2] and mixed with ZnO:Mg nanoparticles [6]. The mechanism for the formation of the excited states in PDPV:Ru includes the ruthenium complex in a stepwise electron transfer process, because the band gap of the ruthenium complex is about 0.5 eV lower than that of the polymer. In the case of the BEHP-co-MEH-PPV-ZnO composite the band gap of ZnO is higher than that of the polymer, and the above mentioned mechanism of the electric-field controlled shift between the blue-green and red emission can not be applied. The electric field-induced switching from heterojunction to bulk charge recombination in bilayer polymer LEDs has been explained recently by the "exciplex" model [7,8]. The formation of "exciplex" states in the polymer-inorganic nanoparticle composites has not been observed in previous studies [2,6]. Fig. 5a shows the positions of the molecular orbitals and the band structure of the BEHP-co-MEH-PPV-ZnO composite components. As can be seen from this diagram, the mechanism for the formation of the excited states in the BEHP-co-MEH-PPV-ZnO composite implies the presence of LUMO-HOMO recombination channel in the BEHP-co-MEH-PPV co-polymer, as well as the contribution to the PL blue emission from radiative deep level recombination in ZnO nanoparticles. It is evident from Fig. 2b, 4 and the inset to Fig. 4, that the former channel can be suppressed by increasing of ZnO concentration and by application of a modest electric field, whereas the latter channel does not demonstrate this influence. The observed in our experiments an additional PL lines at 420 and 460 nm can originate from polymer-nanoparticle charge transfer (CT) complexes formation. As can be seen from Fig. 2a, the PL line at 550 nm corresponds to the excimer emission from original (without complexes with ZnO) co-polymer aromatic components. The influence of electric field on the PL maxima intensity at 420 and 460 nm (as shown in the inset to Fig. 4) indicates the presence of an efficient charge transfer between co-polymer aromatic components (with an energy difference between their HOMO levels of ~ 0.3 eV) and ZnO. We attribute the strong electric-field induced quenching of the PL line at 550 nm in the BEHP-co-MEH-PPV-ZnO composite to the field dependent excimer PL as well as to the partly formation of co-polymer aromatic components – ZnO complexes that also depend on the electric field. In our opinion, an appearance of additional PL lines at 420 and 460 nm (Fig. 2b) in the BEHP-co-MEH-PPV-ZnO composite can be attributed to the "exciplex" - CT - states formation in such polymer-inorganic structure. This conclusion supports by several arguments. First, the energy difference between these PL lines ~ 0.3 eV is similar to that in the pristine co-polymer. Second, there is quenching of the PL line at 650 nm due to the CT complex formation between polymer diphenyl group and ZnO. This process is energetically more favorable when the CT complex formation between benzene fragments and ZnO, and thus it results in quenching the PL line at 650 nm. The capture – recombination mechanism in the BEHP-co-MEH-PPV-ZnO composite consists of two stages: a) charge carriers capture and the "exciplex" state formation; b) radiative recombination of the "exciplex" – CT complex. The latter mechanism can be described by the 3D Onsager model related to the charge carrier photogeneration in organic solids with low charge carrier mobility [13,14]. The model describes the electron-hole pair dissociation under the influence of phonons and electric field as well as the field-assisted thermal dissociation probability, $f_d$, for a pair of charges after application of an electric field. According to this model, $f_d$ determines by the depth of the Coulomb well, which depends on the band gap width (i.e. the ionization potential of donor and electron affinity of acceptor), electric field strength ($\varepsilon$) and the distance (r) between Coulomb centers, which depends on the CT complexes concentration (Fig. 5b). The PL quantum yield in such structure is $I_{PL} = 1 - \eta$, where $\eta$ has the form [15]:

$$\eta(\varepsilon, T) = \varepsilon^n \quad (2)$$

where n ~ 2 [14]. An estimated the quantum yield for CT states in the BEHP-co-MEH-PPV-ZnO composite n ~ 2 at $\varepsilon$ ~ 2 $10^4$ V/cm correlates well with Eq. (2). In addition, the observed fact that an increase in the ZnO nanoparticles concentration influences the recombination channel by quenching green-yellow PL can also be related to the charge transfer between ZnO nanoparticles and BEHP-co-MEH-PPV co-polymer, similar to that reported recently for polymer-ZnMgO-ZnO hybrid nanostructures [16]. As can be seen from Fig. 5b, an increase of the ZnO concentration (i.e. decrease of r), results in the decrease of the PL intensity related to "exciplex"- CT states (420, 460 nm) due to dramatic improvement of charge transfer conditions. Therefore the proposed radiative recombination mechanism based on the 3D Onsager model verified by concentration and electric field dependencies of the PL observed in our composite films. A more detailed study of charge transfer in BEHP-co-MEH-PPV-ZnO composite films within a wide temperature range is needed to clarify an exact recombination mechanism in such structures.

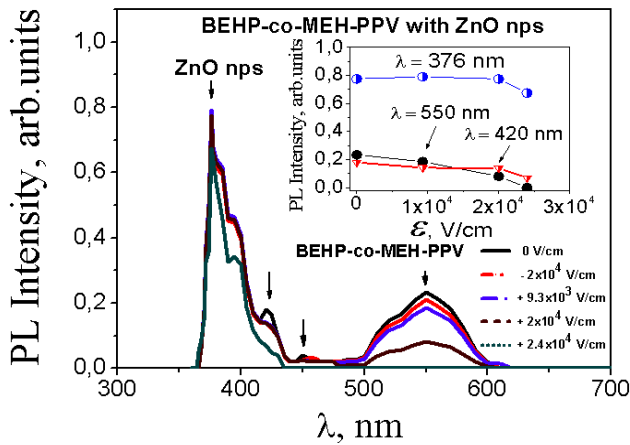

Fig. 4. PL spectrum of the BEHP-*co*-MEH-PPV-ZnO (concentration ratio ~ 1:4) composite film at different electric fields (up to ~ 2.4 10$^4$ V/cm) applied to the planar gold electrodes. Inset shows the influence of electric field on the PL maximum intensity at 550, 420 and 376 nm.

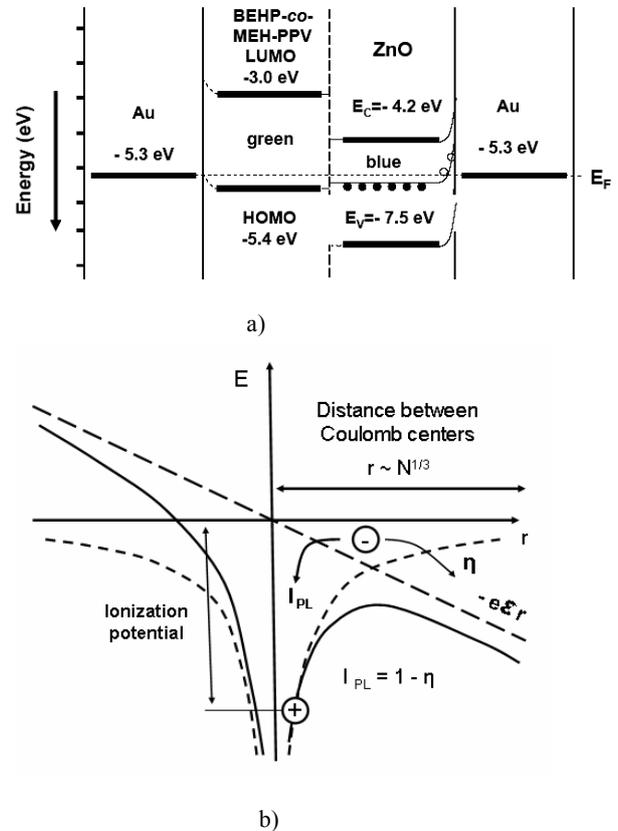

Fig. 5. Band diagram of the Au-BEHP-*co*-MEH-PPV-ZnO-Au composite structure (a); schematic of the CT complex dissociation (b).

**Conclusions**

We report on the effect of electric field and ZnO concentration on the PL from a conjugated polymer BEHP-co-MEH-PPV, mixed with ZnO nanoparticles deposited on a Si-SiO$_2$ substrate with planar Au electrodes. It has been found that the I-Vs of the Au-BEHP-co-MEH-PPV-ZnO-Au composite structure exhibit the non-linear, asymmetric behavior. The PL emission of the BEHP-co-MEH-PPV-ZnO composite has two maxima in the blue and green-yellow regions. Application of a positive voltage bias suppresses the green-yellow emission and shifts the only PL emission maximum towards the blue region. We found that generation of excited states in the BEHP-co-MEH-PPV-ZnO structure includes the formation of "exciplex" states and charge transfer from the polymer to nanoparticles that can be controlled by an electric field. The proposed radiative recombination mechanism based on the 3D Onsager model verified by concentration and electric field dependencies of the PL. There is good reason to believe that the observed effects of electric field and ZnO concentration on the PL of the BEHP-co-MEH-PPV-ZnO composite makes this material promising for making of active layers for multicolor organic displays and white emitters.

**Acknowledgements**

This work supported by Russian Foundation for Basic Research (grant № 07-03-00215) and by Subprogram of Presidium of Russian Academy of Sciences "Polyfunctional materials for molecular electronics".